\documentclass[10pt,conference,hidelinks]{IEEEtran}\IEEEoverridecommandlockouts
\usepackage{epsfig,graphicx,subfigure,psfrag,amsmath,cases}
\usepackage{latexsym,amssymb,amsmath,epsfig,subfigure,algorithm,mathtools}
\usepackage{algorithmic}
\usepackage{color}
\usepackage{url}
\usepackage{scrtime}
\usepackage{cite}
\usepackage{epstopdf}
\usepackage{hyperref}

\author{\authorblockN{Zhiqiang Wei, Derrick Wing Kwan Ng, and Jinhong Yuan}
School of Electrical Engineering and Telecommunications, The University
of New South Wales, Australia \thanks{This work was supported in part by the Australian Research Council (ARC) Linkage Project LP
160100708.}}

\title{Power-Efficient Resource Allocation for MC-NOMA with Statistical Channel State Information}

\newtheorem{T-Prob}{Transformed Problem}

\DeclareMathOperator{\mino}{minimize}

\newcommand{\abs}[1]{\lvert#1\rvert}

\begin{document}
\maketitle

\begin{abstract}
In this paper, we study the power-efficient resource allocation for multicarrier non-orthogonal multiple access (MC-NOMA) systems. The resource allocation algorithm design is formulated as a non-convex optimization problem which takes into account the statistical channel state information at transmitter and quality of service (QoS)  constraints.
 To  strike  a  balance  between  system  performance
and  computational  complexity, we propose a suboptimal power allocation and user scheduling with low computational complexity to minimize the total power consumption. The proposed design exploits the heterogeneity of QoS requirement to determine the  successive interference cancellation decoding order. Simulation results demonstrate that the proposed scheme achieves a close-to-optimal performance and significantly outperforms a conventional orthogonal multiple access (OMA) scheme.
\end{abstract}
\renewcommand{\baselinestretch}{0.95}
\normalsize

\section{Introduction}
Non-orthogonal multiple access (NOMA) has been recognized as a promising multiple access technique for the fifth-generation (5G) wireless networks due to its high spectral efficiency and user fairness \cite{Ding2015b}.
Compared to conventional orthogonal multiple access (OMA), NOMA transmission allows multiple users to share the same frequency resource via exploiting the power domain multiplexing and performing successive interference cancellation (SIC) at the receiver side. It has been shown that NOMA offers considerable performance gains over OMA in previous works \cite{Zhang2016SSR,Liu2016,CN:NOMA_Yan_Sun,Dai2015,Saito2013,Benjebbour2013,Kim2013}. In particular, resource allocation of NOMA has received significant attention since it is critical for the performance of NOMA.  In \cite{Saito2013,Benjebbour2013}, the authors evaluated the system-level performance of NOMA systems. The authors in \cite{Kim2013} studied a minimum total transmission power beamforming problem. In \cite{Sun2015b}, the optimal resource allocation for multiple-input multiple-output (MIMO) NOMA systems to maximize the instantaneous sum-rate was proposed. However, existing works \cite{Saito2013,Benjebbour2013,Kim2013,Sun2015b} on resource allocation of NOMA have relied on the assumption of perfect channel state information at transmitter (CSIT) which is difficult to obtain in practice.

Recently, the notion of imperfect CSIT in NOMA systems for resource allocation algorithm design has been pursued in \cite{Ding2014,Yang2016,Dingtobepublished,Sun2015,Timotheou2015} under various system performance metrics. In \cite{Ding2014}, for a fixed power allocation, the outage probability and ergodic sum-rate of NOMA under statistical CSIT were investigated in a cellular downlink scenario with randomly deployed users. In \cite{Yang2016}, the authors analyzed the performance degradation on these two system performance metrics due to partial CSIT. The authors in \cite{Dingtobepublished} investigated the impact of user pairing on the sum-rate of NOMA for a fixed power allocation scheme. Power allocation was proposed for the maximization of the ergodic capacity and the minimization of the maximum outage probability in \cite{Sun2015} and \cite{Timotheou2015} under statistical CSIT, respectively.

Apart from those performance metrics mentioned above, power efficiency is also important due to the rising energy costs and green communication concerns.  In \cite{Sun2015a}, the authors solved the energy efficiency optimization problem for single-carrier NOMA systems. Yet, if multicarrier NOMA (MC-NOMA) systems are considered, the result from \cite{Sun2015a} may no longer be applicable. In \cite{Di2015,Lei2015,Liu2015}, various power allocation and user scheduling algorithms were proposed to maximize the sum-rate of MC-NOMA systems. However, the results from \cite{Di2015,Lei2015,Liu2015} were based on perfect CSIT assumption which may not available in practice, especially for MC-NOMA systems overloaded with exceedingly number of users. In addition, the aforementioned works have not taken into account the heterogeneous quality of service (QoS) requirements, which play an important role in 5G networks, in particular for small cells and massive access. In fact, power-efficient resource allocation based on statistical CSIT for MC-NOMA systems has not been reported in the literature so far.

In this paper, we focus on the power-efficient resource allocation for MC-NOMA systems with QoS constraints under statistical CSIT. Due to the absence of perfect CSIT, a SIC policy taking consideration of QoS requirements is proposed, where the BS only allows one user to perform SIC. Based on the adopted SIC policy, we formulate the resource allocation problem for MC-NOMA systems to minimize the total transmit power. Since the optimization problem is a mixed combinatorial non-convex problem, a suboptimal solution is proposed to solve the power allocation and the user scheduling problems separately. For a given user scheduling policy, the multicarrier power allocation problem is simplified to a  per-subcarrier basis power allocation problem which facilitates the optimal power allocation design. Interestingly, based on the derived power allocation solution, an explicit metric for SIC decoding order associated with the level of QoS stringency is obtained, as an analogous to the channel gain based SIC decoding order for NOMA with perfect CSIT \cite{Saito2013,Benjebbour2013}. In addition, we have quantified the performance gain of NOMA over OMA in terms of power reduction and shown that the gain increases when the multiplexed users have more distinctive QoS stringency levels. For the user scheduling problem, a low computational complexity suboptimal scheduling algorithm based on agglomerative hierarchical clustering is proposed, which can achieve a close-to-optimal performance. Simulation results show that the proposed scheme significantly increases the power efficiency compared to a conventional OMA scheme.

The rest of the paper is organized as follows. Section II presents the system model and discusses the SIC policy with statistical CSIT. In Section III, we formulate the resource allocation as a non-convex optimization problem with QoS constraints. Section IV and Section V present the solution for power allocation and user scheduling, respectively. Simulation results are presented and analyzed in Section VI. Finally, Section VII concludes this paper.

Notations used in this paper are as follows. Boldface lower case letters denote vectors. $\mathbb{C}$ denotes the set of complex value; $\mathbb{R}^{M\times 1}$ denotes the set of all $M\times 1$ vectors with real entries; $\mathbb{Z}^{M\times 1}$ denotes the set of all $M\times 1$ vectors with integer entries; $\abs{\cdot}$ denotes the absolute value of a complex scalar; $\Pr \left\{  \cdot  \right\}$ denotes the probability of a random event. The circularly symmetric complex Gaussian distribution with mean $\mu$ and variance $\sigma^2$ is denoted by ${\cal CN}(\mu,\sigma^2)$; the uniform distribution in the interval $[a, b]$ is denoted by $U[a,b]$; and $\sim$ stands for ``distributed as".
\begin{figure}[t]
\centering
\includegraphics[width=2.5in]{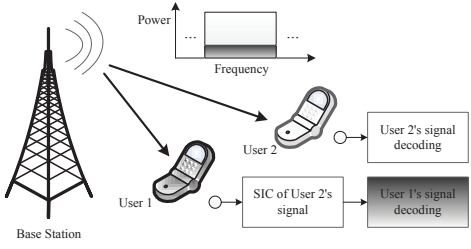}
\caption{A multicarrier downlink NOMA system where two users are multiplexed on one subcarrier in NOMA with perfect CSIT \cite{Benjebbour2013a}. User $1$ has a better channel quality who performs SIC to decode and remove the signal of user $2$ before decoding its desired signal. The allocated power for user $2$ is higher than user $1$.}\label{NOMA_model}
\end{figure}

\section{System Model}
In this section, we present the system model and the adopted assumptions for the considered MC-NOMA system.

\subsection{Multicarrier NOMA System}
A multicarrier downlink NOMA system with one base station (BS) and $K$ downlink users is considered, cf. Figure \ref{NOMA_model}. All transceivers are equipped with a single-antenna. $M$ subcarriers are provided to serve the $K$ users. In this paper, to provide fairness in resource allocation, we assume that only $L$ subcarriers are allocated to one user. In addition, we assume that each of the $M$ orthogonal subcarriers is allocated to at most two users to reduce the computational complexity and delay incurred at receiver side due to the SIC decoding, i.e., $KL \le 2M$. According to the NOMA protocol \cite{Benjebbour2013}, on subcarrier $m \in \left\{ {1, \ldots ,M} \right\}$, the BS transmits the messages of user $i$ and $j$, i.e., $s^m_i$ and $s^m_j$, with transmit power $p^m_i$ and $p^m_j$, $i,j \in \left\{ {1, \ldots ,K} \right\}$, respectively. The corresponding transmitted signal is represented by
\begin{equation}\label{SystemModelTx}
x^m = \sqrt {{p^m_i}} {s^m_i} + \sqrt {{p^m_j}} {s^m_j}.
\end{equation}

The received signal at user $i \in \left\{ {1, \ldots ,K} \right\}$ on subcarrier $m$ is given by
\begin{equation}\label{SystemModelRx}
y_i^m = h_i^m{x^m} + {z_i^m},
\end{equation}
where $z_i^m\sim{\cal CN}(0,\sigma^2)$ denotes the additive white Gaussian noise (AWGN) on subcarrier $m$ at user $i$. Variable $h_i^m \in \mathbb{C}$ represents the channel coefficient including the joint effect of large scale fading and small scale fading, i.e., $h_i^m = \frac{{g_i^m}}{{\sqrt {1 + d_i^\alpha } }}$ and $g_i^m \sim \mathcal{CN}(0,1)$, with $d_i$ denoting the distance between user $i$ to the BS and $\alpha$ denoting the path loss exponent. We assume that the channel gain of small scale fading is Rayleigh distributed and the path loss information is known at the BS due to long term measurement. The cumulative distribution function (CDF) of channel gain of user $i$ on subcarrier $m$ is given by
\begin{equation}\label{ChannelGainCDF}
{F_{{{\left| {{h_i^m}} \right|}^2}}}\left( x \right) = 1 - {e^{ - \left( {1 + d_i^\alpha } \right)x}},\;x \ge 0.
\end{equation}

\subsection{Successive Interference Cancellation Policy}

NOMA exploits the power domain to perform multiple access \cite{Ding2014,Choi2015,Liu2016}. Based on the availability of CSIT, the BS performs user scheduling and power allocation. Besides, SIC will be performed at some of the downlink users to mitigate multi-user interference, cf. Figure \ref{NOMA_model}. In the literature \cite{Saito2013,Benjebbour2013a,Kim2013,Sun2015b}, with perfect CSIT and without QoS consideration, the user with better channel quality (\emph{strong user}) decodes and removes message of user (\emph{weak user}) before decoding its own, while the weak user directly decodes its own message by treating the signal from the strong user as noise. Furthermore, the BS will allocate more power to the weak user to obtain fairness and facilitate SIC process.

Unfortunately, without perfect CSIT, the BS cannot decide the SIC decoding order based on the ordered channel gain information. Similar to the case of NOMA with perfect CSIT, distance might be a criterion to define a strong or weak user. However, this criterion does not take consideration of QoS requirements, which also affect the SIC decoding order and hence change the behavior of power allocation. For example, if a user near the BS requires a lower outage probability, selecting this user to perform SIC needs more transmit power than selecting the other user, which is in contrast to the case of conventional NOMA without considering the QoS requirements.

On the other hand, it can be shown that for the case of NOMA with perfect CSIT and QoS requirements, both users performing SIC will require more transmit power than selecting only the strong user to perform SIC. Intuitively, the allocated power for both users will be increased to cope with the interference in decoding the other user's message. Inspired by this fact, we assume that the BS only allows one user to perform SIC on each subcarrier. Specifically, the BS will select user $i$ to perform SIC on subcarrier $m$ if the power consumption based on this selection is lower than that of selecting user $j$ to perform SIC. Latter in this paper, based on this assumption, an explicit SIC decoding order is derived with the level of QoS stringency.

In addition, given the total required target rate for user $j$ as ${R}_j$, we can split it into $L$ allocated subcarriers equally, since only statistical CSIT is available at the BS. Therefore, the target rate of user $j$ on its allocated subcarrier $m$ is given by
\begin{equation}\label{AssignedRate}
\widetilde{R}_j^m = \frac{{{R_j}}}{L}
\end{equation}
and the corresponding target SINR is given by
\begin{equation}\label{AssignedSINR}
\widetilde{\gamma}_j^m = {2^{\widetilde{R}_j^m}} - 1.
\end{equation}

We assume that SIC at user $i$ on subcarrier $m$ is successful when the achievable rate for decoding the message of user $j$ is not smaller than the target rate of user $j$ on subcarrier $m$, i.e.,
\begin{equation}\label{SICSuccess}
R_{i \to j}^m \ge \widetilde{R}_j^m,
\end{equation}
where $R_{i \to j}^m$ denotes the achievable rate for user $i$ to decode the message of user $j$ on subcarrier $m$ and it is given by
\begin{equation}\label{SICRate}
R_{i \to j}^m = {\log _2}\left( {1 + \frac{{p_j^m{{\left| {h_i^m} \right|}^2}}}{{p_i^m{{\left| {h_i^m} \right|}^2} + \sigma^2}}} \right).
\end{equation}

\section{Problem Formulation}
In this section, we first define the QoS requirements and then formulate the power allocation and user scheduling problem for NOMA systems.

\subsection{Quality of Service}
QoS is usually defined by a target rate and a required outage probability. Given the target rate $\widetilde{R}_i^m$ for each user on each allocated subcarrier, the QoS required by user $i$ on subcarrier $m$ is given by the following outage probability constraint:
\begin{equation}\label{QoSConstraint}
{\mathrm{Pr}}\left\{ {{R_i^m} \ge {{\widetilde R}_i^m}} \right\} \ge 1 - {\delta _i^m},\;\forall i,
\end{equation}
with
\begin{equation}\label{AchievableRate}
R_i^m  = \left\{ {\begin{array}{*{20}{l}}
R_{i,i}^m & \;\mathrm{if}\;{R_{i \to j}^m \ge \widetilde R_j^m},\\
R_{i,j}^m & \;\mathrm{otherwise},
\end{array}} \right.
\end{equation}
where $R_i^m$ and $\delta_i^m$ denote the achievable rate and the required outage probability of user $i$ on subcarrier $m$, respectively. Note that outage probability is defined on each subcarrier, which is commonly adopted in the literature for the simplification of resource allocation design \cite{Zhu2009,Kwan_AF_2010}. Variables $R_{i,i}^m$ and $R_{i,j}^m$ denote the achievable rates for user $i$ on subcarrier $m$ with and without SIC, respectively, and they are given by
\begin{eqnarray}
&& R_{i,i}^m = {\log _2}\left( {1 + \frac{{p_i^m{{\left| {h_i^m} \right|}^2}}}{{\sigma^2}}} \right)\;\mathrm{and}\label{RatesSIC1}\\
&& R_{i,j}^m = {\log _2}\left( {1 + \frac{{p_i^m{{\left| {h_i^m} \right|}^2}}}{{p_j^m{{\left| {h_i^m} \right|}^2} + \sigma^2}}} \right),\label{RatesSIC2}
\end{eqnarray}
respectively.

\subsection{Optimization Problem Formulation}
Now, the joint power allocation and user scheduling design for the MC-NOMA system can be formulated as the following optimization problem:
\begin{subequations}\label{P1}
\begin{equation}\label{P1_Objective}
\hspace*{-2mm}\mathop {\mino }\limits_{{\mathbf{p}},{\mathbf{c}}} \sum\limits_{m = 1}^M {\sum\limits_{i = 1}^K {\sum\limits_{j = 1}^K {c_{i,j}^m\left( {p_i^m + p_j^m} \right)} } }
\end{equation}

\begin{alignat}{1}
\mbox{s.t.}\;\;&{\mathrm{Pr}}\left\{ {{R_i^m} \ge {{\widetilde R}_i^m}} \right\} \ge  c_{i,j}^m\left( 1 - {\delta _i^m} \right),\;\forall i, m,\label{P1_Constraint1}\\
&p_i^m \ge 0,\;\forall i,m,\label{P1_Constraint2}\\
&\sum\limits_{i = 1}^K {\sum\limits_{j = 1}^K {c_{i,j}^m {\leq 2}} } ,\;\forall m,\label{P1_Constraint3}\\
&\sum\limits_{m = 1}^M {\sum\limits_{j = 1}^K {c_{i,j}^m = L} } ,\;\forall i,\label{P1_Constraint4}\\
&c_{i,j}^m \in \left\{ {0,1} \right\},\;\forall i,j,m,\label{P1_Constraint6}
\end{alignat}
\end{subequations}
where $c_{i,j}^m$ is the subcarrier allocation variable which is one if both user $i$ and user $j$ are multiplexed on subcarrier $m$, and will be zero otherwise. Vectors $\mathbf{p}\in\mathbb{R}^{MK \times1}$ and $\mathbf{c}\in\mathbb{Z}^{MK^2 \times1}$ denote the collection of power allocation variables and user scheduling variables. Constraint \eqref{P1_Constraint1} guarantees the QoS of all the users on their allocated subcarriers and it is inactive when user $i$ is not allocated on subcarrier $m$, i.e., $c_{i,j}^m=0$. Constraint \eqref{P1_Constraint2} is non-negative constraint for power allocation variables. Constraints \eqref{P1_Constraint3} and \eqref{P1_Constraint6} are imposed to ensure that at most two users are multiplexed on one subcarrier and all the subcarriers are allocated to reduce the total power consumption. Constraint \eqref{P1_Constraint4} is introduced for resource allocation fairness such that all the users have the same amount of frequency resources.

We note that, for the case of $c_{i,j}^m=1$, $i=j$, subcarrier $m$ is exclusively allocated to user $i$, and the user scheduling policy for subcarrier $m$ is degenerated to conventional orthogonal assignment. In other words, the proposed optimization framework in \eqref{P1} generalizes the resource allocation for conventional OMA as a subcase. We note that the problem in \eqref{P1} is a mixed combinatorial non-convex problem, and there is no systematic and computational efficient approach to solve it optimally. According to \eqref{P1}, the user scheduling is jointly affected by distance $d_i$, target rate ${\widetilde{R}_i^m}$, and required outage probability $\delta_i$, while the counterpart of traditional NOMA with perfect CSIT depends only on channel gain order \cite{Dingtobepublished}. In addition, according to \eqref{P1_Objective} and \eqref{P1_Constraint1}, the power allocation and user scheduling variables are coupled. Therefore, in the following two sections, we propose a suboptimal solution which intends to solve the power allocation and user scheduling separately.

\section{Solution for Power Allocation Problem}\label{Section4}
For a given user scheduling policy $\mathbf{c}$ that satisfies constraints \eqref{P1_Constraint3}, \eqref{P1_Constraint4}, and \eqref{P1_Constraint6}, power allocation can be performed independently on each subcarrier. Therefore, the original problem \eqref{P1} can be simplified to a per-subcarrier two-user power allocation problem. For notational simplicity, we drop the subcarrier index $m$. The simplified optimization problem is given by
\begin{subequations}\label{P2}
\begin{equation}\label{P2_Objective}
\hspace*{-5mm}\mathop {\mino }\limits_{{p_1,p_2}} \;\sum\limits_{i = 1}^2 {{p_i}}
\end{equation}
\vspace*{-5mm}
\begin{alignat}{1}
\mbox{s.t.}\;\;\;\;&{\mathrm{Pr}}\left\{ {{R_i} \ge {{\widetilde R}_i}} \right\} \ge 1 - {\delta _i},\;i \in \left\{ {1,2} \right\},\label{P2_Constraint1}\\
&{p_i} \ge 0,\;i \in \left\{ {1,2} \right\},\label{P2_Constraint2}
\end{alignat}
\end{subequations}
where $R_i$ is given by \eqref{AchievableRate}.

In the following, we first solve the problem in \eqref{P2} by assuming only one user to perform SIC, then derive the SIC decoding order based on power allocation solution, and compare its performance with OMA.

\subsection{Power Allocation Solution}
The optimal power allocation solution for the problem in \eqref{P2} can be obtained via the following two cases. For the first case, we only allow user $1$ to perform SIC and obtain the corresponding power allocation solution. The power allocation solution for the second case, which only allows user $2$ to perform SIC, is also obtained. Then, the optimal solution for the problem in \eqref{P2} is given by the solutions for both cases with the lower power consumption.

According to \eqref{SICSuccess}, if we allow user $1$ to perform SIC and prevent user $2$ to do that, the following prerequisites should be satisfied:
\begin{eqnarray}
&&{p_2} - {p_1}{\widetilde{\gamma} _2} > 0\quad\mathrm{and}\label{Prerequisite1}\\
&&{p_1} - {p_2}{\widetilde{\gamma} _1} \le 0.\label{Prerequisite2}
\end{eqnarray}

We note that, due to the channel uncertainty, the prerequisite in \eqref{Prerequisite1} cannot guarantee the success of SIC and the success of SIC also cannot guarantee outage free transmission. This makes the resource allocation for MC-NOMA in this paper fundamentally different from the case of perfect CSIT. Under these two prerequisites, \eqref{Prerequisite1} and \eqref{Prerequisite2}, the outage probability for both users are given by
\begin{eqnarray}
&&{\rm{P}}^{\mathrm{out}}_1 = {\rm{Pr}}\left\{ {{R_{1 \to 2}} \ge {{\tilde R}_2},{R_{1,1}} < {{\tilde R}_1}} \right\}\notag\\
&&\hspace*{10mm}+ {\rm{Pr}}\left\{ {{R_{1 \to 2}} < {{\tilde R}_2},{R_{1,2}} < {{\tilde R}_1}} \right\},\label{OutageProbability1}\\
&&{\rm{P}}^{\mathrm{out}}_2 = {\rm{Pr}}\left\{ {{R_{2,1}} < {{\tilde R}_2}} \right\},\label{OutageProbability2}
\end{eqnarray}
where ${\rm{P}}^{\mathrm{out}}_1$ and ${\rm{P}}^{\mathrm{out}}_2$ denote the outage probability of user $1$ and user $2$, respectively. Equation \eqref{P2_Constraint1} requires that ${\rm{P}}^{\mathrm{out}}_i \le \delta _i$. Note that ${\rm{P}}^{\mathrm{out}}_1$ consists of two terms which denote the outage probability with a successful SIC and an unsuccessful SIC at user $1$, respectively.

Substituting \eqref{SICRate}, \eqref{RatesSIC1}, and \eqref{RatesSIC2} into \eqref{OutageProbability1} and \eqref{OutageProbability2} yields
\begin{eqnarray}
&&\hspace*{-12mm}{\rm{P}}^{\mathrm{out}}_1 = {\rm{Pr}}\left\{ {{{\left| {{h_1}} \right|}^2} < \max \left( {\frac{{{\widetilde{\gamma} _1}{\sigma ^2}}}{{{p_1}}},\frac{{{\widetilde{\gamma} _2}{\sigma ^2}}}{{ {{p_2} - {p_1}{\widetilde{\gamma} _2}} }}} \right)} \right\} \; \mathrm{and} \label{OutageProbability3}\\
&&\hspace*{-12mm}{\rm{P}}^{\mathrm{out}}_2 = {\rm{Pr}}\left\{ {{{\left| {{h_2}} \right|}^2} < \frac{{{\widetilde{\gamma} _2}{\sigma ^2}}}{{ {{p_2} - {p_1}{\widetilde{\gamma} _2}} }}} \right\},\label{OutageProbability4}
\end{eqnarray}
respectively.

 Exploiting the CDF of channel gain \eqref{ChannelGainCDF} in \eqref{OutageProbability3} and \eqref{OutageProbability4}, and substituting them into \eqref{P2_Constraint1}, we obtain the solution of \eqref{P2} with the minimized total transmit power as:
\begin{eqnarray}
&&\hspace*{-5mm}{p_1^{(1)}} = \frac{{{\widetilde{\gamma} _1}}}{{{\beta _1}}}\quad \mathrm{and}\label{Solution1}\\
&&\hspace*{-5mm}{p_2^{(1)}} = \max \left( {\frac{{{\widetilde{\gamma} _1}{\widetilde{\gamma} _2}}}{{{\beta _1}}} + \frac{{{\widetilde{\gamma} _2}}}{{{\beta _1}}},\frac{{{\widetilde{\gamma} _1}{\widetilde{\gamma} _2}}}{{{\beta _1}}} + \frac{{{\widetilde{\gamma} _2}}}{{{\beta _2}}},\frac{1}{{{\beta _1}}}} \right),\label{Solution2}
\end{eqnarray}
where ${\beta _i} =  - \frac{{\ln \left( {1 - {\delta _i}} \right)}}{{\sigma^2\left( {1 + d_i^\alpha } \right)}}$, ${p_i^{(1)}}$, $i \in \left\{ {1,2} \right\}$, is the allocated power for user $i$ for the first case, and the superscript $(1)$ denotes allowing user $1$ to perform SIC. Note that $\frac{1}{{{\beta _i}}}$ can be interpreted as the level of QoS stringency for user $i$, where a large $\frac{1}{{{\beta _i}}}$ means user $i$ is far away from the BS or has a small required outage probability, such that a higher transmit power is necessary to satisfy its stringent QoS requirement. Similar to the case of NOMA with perfect CSIT, we can define a user with larger ${{{\beta _i}}}$ as a \emph{QoS non-demanding user} and define the other user as a \emph{QoS demanding user}.

For the second case which allows user $2$ to perform SIC and prevents user $1$ to do that, the prerequisites are given by
\begin{equation}\label{Prerequisite34}
{p_1} - {p_2}{\widetilde{\gamma} _1} > 0 \quad \mathrm{and}\quad
{p_2} - {p_1}{\widetilde{\gamma} _2} \le 0.
\end{equation}

Similarly, the power allocation solution for \eqref{P2} can be derived and given as
\begin{eqnarray}
&&\hspace*{-5mm}{p_1^{(2)}} = \max \left( {\frac{{{\widetilde{\gamma} _1}{\widetilde{\gamma} _2}}}{{{\beta _2}}} + \frac{{{\widetilde{\gamma} _1}}}{{{\beta _2}}},\frac{{{\widetilde{\gamma} _1}{\widetilde{\gamma} _2}}}{{{\beta _2}}} + \frac{{{\widetilde{\gamma} _1}}}{{{\beta _1}}},\frac{1}{{{\beta _2}}}} \right)\;\mathrm{and}\label{Solution3}\\
&&\hspace*{-5mm}{p_2^{(2)}} = \frac{{{\widetilde{\gamma} _2}}}{{{\beta _2}}},\label{Solution4}
\end{eqnarray}
where the superscript $(2)$ denotes allowing user $2$ to perform SIC.

In summary, the optimal solution for the problem in \eqref{P2} can be selected by
\begin{equation}\label{GlobalSolution}
\left( {{{p_1}},{{p_2}}} \right) =
\left\{ \begin{array}{ll}
\hspace*{-1mm}\left( p_1^{(1)}, p_2^{(1)}\right)& \mathrm{if} \; p_1^{(1)}\hspace*{-1mm}+\hspace*{-1mm}p_2^{(1)} \le p_1^{(2)}\hspace*{-1mm}+\hspace*{-1mm}p_2^{(2)}, \\
\hspace*{-1mm}\left( p_1^{(2)}, p_2^{(2)}\right)& \mathrm{otherwise},
\end{array} \right.
\end{equation}
and the BS will inform user $i$ to perform SIC and forbid the other user to do that if $\left( p_1^{(i)}, p_2^{(i)}\right)$ is selected.

\subsection{A Simple SIC Decoding Order}
The selection of optimal power allocation in \eqref{GlobalSolution} incorporates the SIC decoding policy implicitly to achieve minimum power consumption. However, we can obtain an explicit rule to determine the SIC decoding order for a general condition of ${\widetilde{\gamma} _1}\ge1$ and ${\widetilde{\gamma} _2}\ge1$, which means that both users' target rates are not smaller than 1 bit/s/Hz. In such a condition, $\frac{1}{{{\beta _1}}}$ and $\frac{1}{{{\beta _2}}}$ will never be chosen in \eqref{Solution2} and \eqref{Solution3}, respectively. Thus, we have a simple solution for the joint optimal SIC decoding order and power allocation:
\begin{equation}\label{SICPolicy}
\left( {{{p_1}},{{p_2}}} \right) =
\left\{ \begin{array}{ll}
\left( p_1^{(1)}, p_2^{(1)}\right)& \mathrm{if} \; {{{\beta _1}}} \ge {{{\beta _2}}}, \\
\left( p_1^{(2)}, p_2^{(2)}\right)& \mathrm{otherwise},
\end{array} \right.
\end{equation}
which indicates that the QoS non-demanding user is always selected to perform SIC to minimize the power consumption.

Therefore, for a general condition of ${\widetilde{\gamma} _1}\ge1$ and ${\widetilde{\gamma} _2}\ge1$, ${\beta _i}$ defines the optimal SIC decoding policy in terms of power efficiency, where we only allow the QoS non-demanding user to perform SIC to reduce the total power consumption. Note that for ${\widetilde{\gamma} _i}<1$, we have to evaluate both solutions and compare them in \eqref{GlobalSolution} to find the SIC decoding order.

\subsection{Comparison between NOMA and OMA}
For the case of NOMA with perfect CSIT, it is well known that the performance gain of NOMA over OMA increases when the differences in channel gains between the multiplexed users become larger \cite{Benjebbour2013a,Saito2013,Dingtobepublished}. In this paper, we can obtain a similar conclusion with our scheme in terms of power reduction for the case of imperfect CSIT. For a fair comparison, we impose the same spectral efficiency for NOMA and OMA, where a single subcarrier is further split into two subcarriers with equal bandwidth for the OMA case. Therefore, the power allocation for two OMA users with statistical CSIT on one subcarrier is given by
\begin{equation}\label{OMA1}
\left( {p_1^{\mathrm{OMA}},p_2^{\mathrm{OMA}}} \right)=\left( \frac{{{2^{2{{\widetilde R}_1}}} - 1}}{{{2\beta _1}}},\frac{{{2^{2{{\widetilde R}_2}}} - 1}}{{{2\beta _2}}} \right),
\end{equation}
where the superscript ``$\mathrm{OMA}$" denotes the case of OMA.

Now, we provide a sufficient condition that  the power consumption of NOMA is no larger than that of OMA. Suppose ${\widetilde{R} _1} \ge 1$ bit/s/Hz and ${\widetilde{R} _2} \ge 1$ bit/s/Hz, we can obtain the performance gain of NOMA over OMA in terms of power reduction as follows:
\begin{eqnarray}\label{NOMAOverOMA}
&&\hspace*{-8mm}p_{\mathrm{total}}^{\mathrm{OMA}} - p_{\mathrm{total}}^{\mathrm{NOMA}} = \notag\\
&&\hspace*{-8mm}\left\{ \begin{array}{ll}
\hspace*{-1mm}\frac{{{\widetilde{\gamma} _1}{\widetilde{\gamma} _2}}}{{\sqrt {{\beta _1}} }}\left( {\frac{1}{{\sqrt {{\beta _2}} }} \hspace*{-1mm}-\hspace*{-1mm} \frac{1}{{\sqrt {{\beta _1}} }}} \right) + \frac{1}{2}{\left( {\frac{{{\widetilde{\gamma} _2}}}{{\sqrt {{\beta _2}} }} \hspace*{-1mm}-\hspace*{-1mm} \frac{{{\widetilde{\gamma} _1}}}{{\sqrt {{\beta _1}} }}} \right)^2}\ge 0& \hspace*{-1mm}\mathrm{if} \; {{{\beta _1}}} \ge {{{\beta _2}}}, \\
\hspace*{-1mm}\frac{{{\widetilde{\gamma} _1}{\widetilde{\gamma} _2}}}{{\sqrt {{\beta _2}} }}\left( {\frac{1}{{\sqrt {{\beta _1}} }} \hspace*{-1mm}-\hspace*{-1mm} \frac{1}{{\sqrt {{\beta _2}} }}} \right) + \frac{1}{2}{\left( {\frac{{{\widetilde{\gamma} _1}}}{{\sqrt {{\beta _1}} }} \hspace*{-1mm}-\hspace*{-1mm} \frac{{{\widetilde{\gamma} _2}}}{{\sqrt {{\beta _2}} }}} \right)^2}>0& \hspace*{-1mm}\mathrm{otherwise},
\end{array} \right.
\end{eqnarray}
where $p_{\mathrm{total}}^{\mathrm{OMA}}$ and $p_{\mathrm{total}}^{\mathrm{NOMA}}$ denote the total power consumption of OMA and NOMA on a single subcarrier, respectively. It can be observed that under the sufficient condition, the power reduction of NOMA over OMA is non-negative. More importantly, with ${\widetilde{R} _1} \ge 1$ bit/s/Hz and ${\widetilde{R} _2} \ge 1$ bit/s/Hz, the performance gain of NOMA over OMA also increases when the difference in the level of QoS stringency or the target rate between the QoS demanding user and QoS non-demanding user become larger, e.g. $\beta_1 \gg \beta_2$ or $\widetilde{\gamma} _1 \gg \widetilde{\gamma} _2$. Note that the total power consumption difference between NOMA and OMA is zero for the case of statistical CSIT when the multiplexed users have identical distances and QoS requirements, i.e., $\beta_1 = \beta_2$ and $\widetilde{\gamma}_1=\widetilde{\gamma}_2$.

In summary, from \eqref{SICPolicy} and \eqref{NOMAOverOMA}, we conclude that the level of QoS stringency $\frac{1}{{{\beta _i}}}$ plays a significant role in power allocation and SIC decoding order design.
\section{User Scheduling Algorithm}
According to \eqref{P1}, the $K$ users can be treated as $KL$ independent virtual users since their QoS constraints \eqref{P1_Constraint1} are imposed on each subcarrier independently. In addition, NOMA provides significant system performance gain in high system load scenario. Thus we focus on a practical overload scenario, i.e., $KL>M$. Now, to serve $K$ users via $M$ subcarriers, we intend to generate a user combination consisting of $KL-M$ users pairs and $2M-KL$ single users, which correspond to NOMA and OMA, respectively.
In all candidate user combinations, user scheduling needs to select one combination which consumes the minimal power.

Without loss of generality, we assume $L=1$ in this section. For $L>1$, we simply eliminate the combinations where one user is paired with itself to satisfy constraint \eqref{P1_Constraint4}. As we mentioned before, the power consumption of each subcarrier is only affected by the two multiplexed users on it. To schedule $K$ user over $M$ subcarriers, there are $\dbinom{K}{2}$ possible candidate pairs of users, and we can get the power consumption for all the pairs from \eqref{GlobalSolution}, i.e., $p_{ij}$, $i,j \in \left\{ {1, \ldots ,K} \right\}$ and $i \ne j$. In addition, for orthogonal subcarrier assignment, the power consumption for user $i$ is given by $p_{ii}=\frac{{{\widetilde{\gamma} _i}}}{{{\beta _i}}}$.

The number of all candidate combinations is given by
\begin{equation}\label{TotalNumber}
N = \dbinom{K}{2M-K} \prod\limits_{m = 1}^{K-M} {\left( {2m - 1} \right)}.
\end{equation}
To obtain the optimal user scheduling policy, we need to verify and compare all these $N$ candidate combinations, where $N$ is prohibitively large even for moderate $K$ and $M$. Thus, we attempt to propose a heuristic user scheduling algorithm based on the following geometric illustration.

\begin{figure}[t]
\centering
\includegraphics[width=2.5in]{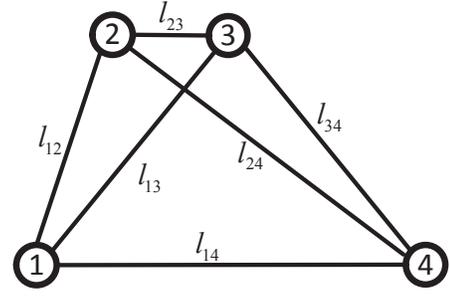}
\caption{Geometric illustration for the proposed user scheduling with $4$ users.}\label{UserScheduling}\vspace*{-2mm}
\end{figure}
\begin{figure}[t]
\centering
\includegraphics[width=3.5in]{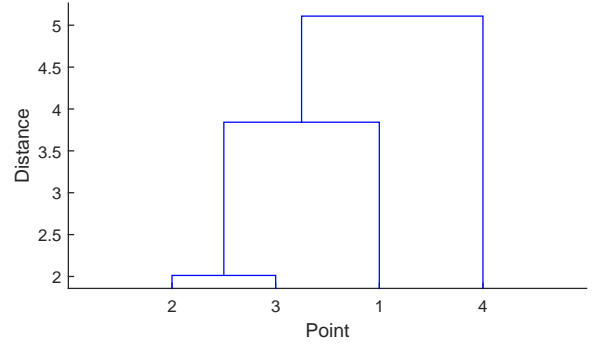}
\caption{Dendrogram for the case in Figure \ref{UserScheduling}.}\label{DendrogramCase}\vspace*{-4mm}
\end{figure}

Figure \ref{UserScheduling} illustrates a user scheduling case with $4$ users, where every point denotes a user, every line $l_{ij}$ denotes pairing user $i$ and user $j$. The length of line $l_{ij}$ is given by $p_{ij}$, which denotes the power consumption for pairing user $i$ and user $j$. Note that it is an undirected graph, i.e., $p_{ij} = p_{ji}$ since both $p_{ij}$ and $p_{ji}$ denote the power consumption for pairing user $i$ and user $j$. From a minimum power consumption perspective, user $i$ and user $j$ are more likely to be paired with each other if point $i$ and point $j$ are close and they are far away from other points, such as point $2$ and point $3$ in Figure \ref{UserScheduling}. Based on this simple idea, the user scheduling problem becomes a clustering problem among all the points on a two-dimensional plane.

Now, we apply agglomerative hierarchical clustering to build the hierarchy from the individual points by progressively merging clusters \cite{Kaufman2009}. Based on the length of all the lines $l_{ij}$, we can obtain the dendrogram structure of all the points, which illustrates the arrangement of clusters, cf. Figure \ref{DendrogramCase}. In the dendrogram, the vertical axis of a point denotes the average distance between this point with all the clusters below it, and the horizon axis presents the ordered points set in terms of distance to the clusters on the left of it. For example, Figure \ref{DendrogramCase} illustrates a dendrogram for the case in Figure \ref{UserScheduling}, where point $3$ is the nearest point to point $2$, and point $4$ is the farthest point to the cluster consists of points $2$, $3$, and $1$. In other words, the horizon axis illustrates the ordered users set in terms of average power consumption for pairing with each user on its left. Note that the generated order in dendrogram is based on the joint effect of $\frac{1}{{{\beta _i}}}$ and $\widetilde{\gamma} _i$, which provides a rule of thumb for user scheduling.

According to \eqref{Solution1}, \eqref{Solution2}, \eqref{Solution3}, and \eqref{Solution4}, the power consumption of NOMA increases with the target SINR of both multiplexed users. Thus we expect that the right $2M-K$ users in the horizontal axis of the dendrogram are assigned on $2M-K$ subcarriers exclusively to reduce the total system power consumption since these users are usually QoS demanding. For the remaining left $2K-2M$ users, we need to generate $K-M$ pairs of users. Since the performance gain over OMA increases with the difference of $\frac{1}{{{\sqrt{\beta _i}}}}$ as well as ${\frac{{{\gamma _i}}}{{\sqrt {{\beta _i}} }}}$ between paired users, referring to \eqref{NOMAOverOMA}, we intend to partition the remaining left $2K-2M$ users into two groups and pair them in successive order. For example, if we only have two subcarriers for the case of $4$ users in Figure \ref{DendrogramCase}, we  partition them into two groups, $\left\{ {2,3} \right\}$ and $\left\{ {1,4} \right\}$. Then we pair user $2$ with user $1$  and pair user $3$ with user $4$. The user scheduling algorithm is summarized in \textbf{Algorithm 1}.

\begin{table}[t]

\begin{algorithm} [H]                    
\caption{User Scheduling Algorithm}     
\label{alg2}                           
\begin{algorithmic} [1]
\small          
\STATE Compute $p_{ij}$, $i \ne j$, $i,j \in \left\{ {1, \ldots ,K} \right\}$, through \eqref{GlobalSolution}.
\STATE Generate the dendrogram based on $p_{ij}$ via agglomerative hierarchical clustering \cite{Kaufman2009}.
\STATE Allocate the right $2M-K$ users on $2M-K$ subcarriers exclusively.
\STATE Partition the left $2K-2M$ users into two groups and pair them in successive order on $K-M$ subcarriers.
\end{algorithmic}
\end{algorithm}\vspace*{-8mm}
\end{table}

Note that for the case of $L>1$, each point in Figure \ref{UserScheduling} will be replaced by $L$ points which have the same distance to all the other points due to our equally target rate assignment \eqref{AssignedRate}. Therefore, the dendrogram in Figure \ref{DendrogramCase} will be extended by replacing each user with a cluster of $L$ users of equal altitude. Therefore, our scheduling will avoid the pair of users where one user is paired with itself unless $KL-M=1$. For the case of $KL-M=1$, we will select the first point of the right $2M-K$ users to pair to satisfy constraint \eqref{P1_Constraint4}.

We note that although the proposed user scheduling algorithm is suboptimal, it is more computational efficient compared to optimal exhaustive search. In particular, the complexity of agglomerative clustering algorithm is only $\mathcal{O}\left(K^3\right)$ in general case. Besides, the suboptimality of the proposed user scheduling algorithm will be verified in the simulation section.

\section{Results}
In this section, the performance of our proposed scheme is verified with simulations. In a single-cell with BS located at the center with cell size $D$, there are $K$ users randomly and uniformly distributed between $30$ m and $D$ m, i.e., $d_i \sim U[30,\;D]$ m. Similarly, the target rates of all the users are generated by $\widetilde{R}_i \sim U[0.1,\;10]$  bit/s/Hz. In the following simulations, two kinds of outage probability are evaluated to compare the performance gain by introducing the QoS constraints: Case I with equal outage probability $\delta_i = 10^{-2}$ and Case II with random outage probability $\delta_i \sim U[10^{-5},\;0.1]$. The user noise power on each subcarrier is $\sigma^2=-128$ dBm. The 3GPP path loss model with path loss exponent $\alpha = 3.6$ is adopted in our simulations \cite{Access2010}.
The simulation results shown in the sequel are averaged over $1000$ realizations of different user distances, target rates, multipath fading coefficients, and outage probability requirements.
\subsection{Power Consumption versus Cell Size}
\begin{figure}[t]
\centering
\includegraphics[width=3.5in]{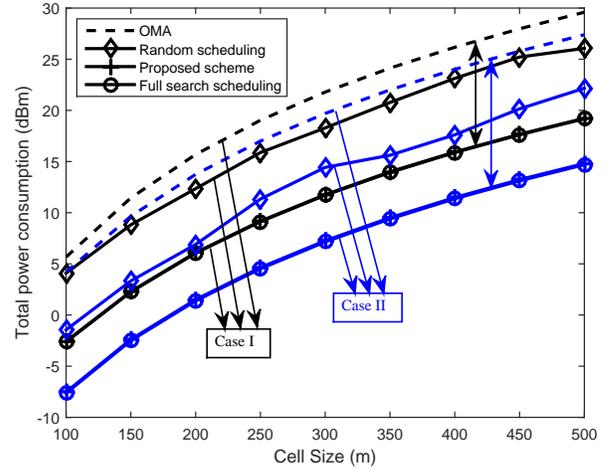}
\caption{Power consumption versus cell size. The results for Case I and Case II are illustrated with black color and blue color, respectively. The double-sided arrows illustrate the performance gain of our proposed scheme over OMA in Case I and Case II, respectively.}
\label{PC_vs_CellSize}
\end{figure}

In Figure \ref{PC_vs_CellSize}, we investigate the power consumption versus cell size $D$ for the considered MC-NOMA system with $M=5$, $K = 4$, and $L=2$ \footnote{Since the computational complexity of full search is extremely large, we adopt small values for $M$, $K$, and $L$ to compare our proposed scheme with the full search scheduling. We note that our proposed scheme is very computational efficient compared to exhaustive search, which can apply to a scenario with more users and subcarriers.}. For comparison, we also show the performance of OMA, random scheduling and, full search scheduling. Note that power consumption for OMA is given with \eqref{OMA1} by replacing $2{{\widetilde R}_i}$ with $K/M{{\widetilde R}_i}$ since the available frequency bandwidth is split equally for $K$ users. In addition, we note that the random scheduling and the full search scheduling are performed together with the proposed power allocation solution \eqref{GlobalSolution}. It can be seen that our proposed user scheduling method provides a significant power saving compared to the random scheduling, and achieves a performance close to the full search scheduling in both cases. The reason for this improvement is that our proposed user scheduling method takes into account the joint effect of $\frac{1}{{{\beta _i}}}$ and $\widetilde{\gamma} _i$, and exploits the heterogeneity of QoS requirements, which achieves a better utilization of power domain. More importantly, the performance gain of our proposed scheme over OMA in Case II is larger than Case I. This result demonstrates the effectiveness of our proposed scheme in exploiting the QoS heterogeneity to reduce power consumption.

\subsection{Total Power Consumption versus Number of Users}
\begin{figure}[t]
\centering
\includegraphics[width=3.5in]{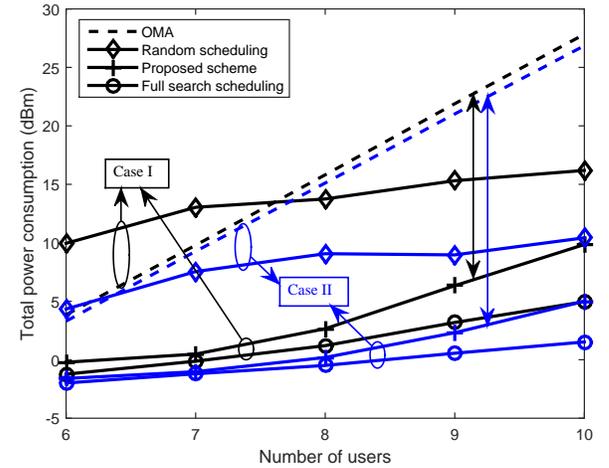}
\caption{Total power consumption versus the number of users. The results for Case I and Case II are illustrated with black color and blue color, respectively. The double-sided arrows illustrate the performance gain of our proposed scheme over OMA in Case I and Case II, respectively.}
\label{UserSchedulingAlgorithm}
\end{figure}
In Figure \ref{UserSchedulingAlgorithm}, we investigate the performance of our proposed scheme versus the number of users. A MC-NOMA system with $M=5$ subcarriers and cell size $D=200$ m is considered. We assume $L = 1$ in this case, and the number of users $K$ varies from $6$ to $10$. It can be observed that our proposed resource allocation scheme reduces the power consumption substantially compared to the random scheduling, and also performs closely to the full search scheduling in both cases. Besides, it can be observed that the performance gain of our proposed scheme over OMA in Case II is also larger than that in Case I owing to a better resource utilization via taking into account the diversification of QoS requirements. More importantly, it can be seen that the performance gain over OMA increases with number of users in both cases, which is consistent with the case of NOMA with perfect CSIT. In fact, the QoS requirements and average user channel gain become more heterogeneous for an increasing number of users. Thus the conventional OMA scheme fails to accommodate the diverse needs, and our proposed scheme will obtain more performance gain.

We note that the power consumption of OMA is lower than that of random scheduling for $K=6$ and $K=7$ but still higher than our proposed scheme. Although NOMA significantly outperforms OMA in single subcarrier \eqref{NOMAOverOMA}, MC-NOMA with random scheduling may consume more power than OMA in such low overload ratio, e.g. $\frac{K}{M}\approx 1$. In fact, with low overload ratio, due to the user scheduling constraint \eqref{P1_Constraint4} for fairness consideration, there are only few pairs of users in the user combination, and thus MC-NOMA with random scheduling cannot fully exploit the power domain.
In addition, we note that the performance gap between our proposed suboptimal scheme with the full search scheduling increases slightly with the number of users. However, the computational complexity of our proposed scheme is significantly lower than that of the full search scheduling.

\section{Conclusion}
In this paper, we studied the power-efficient resource allocation for MC-NOMA systems with statistical CSIT by taking into account the heterogeneity of QoS requirements. The resource allocation was formulated as a non-convex optimization problem to minimize the total power consumption. A low computational complexity suboptimal solution was proposed to solve power allocation and user scheduling problems separately. We derived the power allocation solution and characterized an explicit metric to decide the SIC decoding order associated with the level of QoS stringency. Besides, under a sufficient condition, we showed that the performance gain of NOMA over OMA in terms of power reduction increases with the difference in the level of QoS stringency between the multiplexed users. For the user scheduling, a computational efficient scheduling algorithm based on agglomerative hierarchical clustering was proposed. Simulation results demonstrated that our proposed scheme achieves a close-to-optimal performance and significant outperforms the conventional OMA scheme. Furthermore, our results also showed the effectiveness of our proposed scheme in exploiting the QoS heterogeneity to reduce power consumption.

\bibliographystyle{IEEEtran}
\bibliography{PowerEfficient}

\end{document}